\documentstyle[preprint,prl,aps]{revtex}
\input epsf.tex
\def\simlt{\mathrel{\lower .3ex \rlap{$\sim$}\raise .5ex \hbox{$<$}}}
\def\simgt{\mathrel{\lower .3ex \rlap{$\sim$}\raise .5ex \hbox{$>$}}}

\begin{document}
\draft
\tighten
\title{Self-Organized Short-Term Memories}
\vspace{1cm}

\author{
S.N. Coppersmith\thanks{The James Franck
Institute, The University of Chicago, 5640 Ellis Avenue,
Chicago, IL  60637}
\and
T.C. Jones\thanks{Department of Physics and Astronomy,
Clemson University, Clemson, SC  29694-1911}
\and
L.P. Kadanoff\footnotemark[1]
\and
A. Levine\thanks{Exxon Research \& Engineering Company,
Route 22 East, Annandale, NJ  08801}
\and
J.P. McCarten\footnotemark[2]
\and
S.R. Nagel\footnotemark[1]
\and
S.C. Venkataramani\footnotemark[1]
\and
Xinlei Wu\footnotemark[2]
}
\date{\today}

\maketitle

\baselineskip = 24pt

\begin{abstract}
\baselineskip = 24 pt
We report short-term memory formation in a nonlinear dynamical system with
many degrees of freedom.  The system ``remembers'' a sequence of impulses
for a transient period, but
it coarsens and eventually ``forgets'' nearly all of them.
The memory duration increases as the number of degrees of freedom
in the system increases.
We demonstrate the existence of these transient
memories in a laboratory experiment.
\end{abstract}

\vskip .5 cm
%
%
\newpage


\baselineskip = 24pt
We present a deterministic nonlinear dynamical system with many degrees of
freedom which self-organizes to store memories, in that a
configuration-dependent quantity ``learns" preselected values.  The system,
a simple discretized diffusion equation,
encodes multiple memories during an
extended transient period, but, in the limit of long times, retains no more
than two of them.  This system thus displays a mechanism by which memories
are forgotten as well as learned.

We demonstrate: (1) Short-term memories are exhibited by a system with two
degrees of freedom, $N=2$, and become more pronounced as $N$
is increased.  (2) The interval in which multiple memories are
encoded typically grows as the square of the system's linear extent.
(3) Many features of the dynamics, including their duration, can be
understood analytically.
(4) The mechanism is robust and is manifest in experiments on
a sliding charge-density wave solid.

Consider a system of coupled maps:
\begin{equation}
x_j(\tau+1) \ =\  x_j(\tau) + {\rm int} \left [
k\sum_{i~(nn)}(x_i(\tau)-x_j(\tau))\ + (1-A_\tau) \right ] ,
\label{eq:xeq}
\end{equation}
where $i,j$ are the particle indices, the sum is over nearest neighbors,
$\tau $ is the time index, and ${\rm int} [z]$ is the largest integer less
than or equal to $z$.  These equations
describe the evolution of the
positions $x_j$ of $N$ particles in a deep periodic potential, with nearest
neighbor particles connected by springs of spring constant $k \ll 1$,
in the presence of force impulses $(1-A_\tau)$.
They describe the dynamics of sliding
charge-density waves (CDW's),\cite{cdwrefs,coppersmith87a,coppersmith87b},
and are closely related to models of a variety of dynamical
systems.\cite{othermodels}  The one-dimensional, $\tau$-independent
version of this system ($A_\tau = A$) has been studied
previously\cite{coppersmith87a,tang87}(see also
Ref.~\cite{coppersmith87b}).  Here, we consider $A$'s which repeatedly
cycle through $M$ different values.

The self-organization that occurs as these maps evolve is manifest in the
discrete curvature variables\cite{tang87},
$c_j(\tau) \ =\ k\sum_{i~(nn)}(x_i(\tau)-x_j(\tau))$, which obey:
\begin{equation}
c_j(\tau +1) - c_j(\tau) =
k \sum_{i~(nn)}
\left \{
{\rm int} \left [ c_i(\tau) + (1-A_\tau) \right ]
- {\rm int} \left [ c_j(\tau) + (1-A_\tau) \right ]
\right \} \ .
\label{eq:curveq}
\end{equation}

Figure 1 shows normalized histograms of ${\rm frac}(c) = c-{\rm int}(c)$
for a two-dimensional system with $M=5$, periodic boundary conditions, and
a random initial configuration of $x$'s.  Memory encoding is shown by
the accumulation of $c$'s with ${\rm frac}(c)= {\rm frac}(A_\tau)$.  For a
while all $M$ memories are encoded to a similar degree; eventually all are
forgotten except for two values of A.\cite{footnote4}.
No evolution occurs after the last trace, a fixed point of the map
(\ref{eq:curveq}).

Figure 2 shows the curvature variables $c_j(\tau)$ versus time $\tau $ for
one-dimensional chains with one free and one fixed end:
\begin{equation}
x_0(\tau)=0;~~x_{N+1}(\tau)=x_N(\tau)~~~~~ {\rm for~all}~\tau
\label{eq:fixedendbc}
\end{equation}
During the evolution, each $c_j$ sticks at values corresponding to each
$A_\tau $.  This tendency is more pronounced for $N=10$ than for $N=2$,
indicating that larger systems encode transient memories more effectively.
At the fixed point, only one memory (rather than two as in the model with
periodic boundary conditions) is encoded.

In CDW experiments, memory encoding is manifest
as synchronization of the response
to a repeated train of
driving pulses so that V/I (V=voltage, I=CDW current,
which is proportional to the CDW velocity $v_{CDW}$)
{\rm decreases} just as each pulse ends.
The correspondence between $V/I$
and the $c$'s
is discussed in detail in
Refs.~\cite{coppersmith87a,coppersmith87b}.
Heuristically, it follows because
$x_j(\tau)$ can be thought of as the position of particle $j$
after pulse $\tau$, and the ${\rm int}$ functions in Eqs.~(1) arise
because after
each pulse every particle falls into the nearest potential minimum.
The memory values are at the discontinuities of the
the ${\rm int}$ functions, which
for the highly overdamped dynamics relevant to CDW's\cite{cdwrefs}
means that many particles are at potential
maxima at the end of a pulse.
Since particles mounting the potential go slower than those
descending it (again, implied by overdamped dynamics),
when many particles have $c$'s on memory values, then
a preponderance of particles are at potential maxima, which
in turn implies that the ratio $v_{CDW}/V$ is increasing
at the end of each pulse.
{\em Single} memory retention using identical pulses
has been seen previously.\cite{fleming86,brown86}
Here we report {\em multiple} memory encoding.\cite{expfootnote}
Figure 3 shows the successful training
of a sample using 5 different four-pulse
sequences (current pulses).
For this sample, we investigated 25
different four-pulse sequences and observed that
the voltage response at the
end of each pulse had a negative slope
(indicating the
retention of a memory) 85\% and a positive slope
only 5\% of the time.
Thus, multiple memories are observed
in experiments as well as simulations.

To understand why the memories form, consider the ``nailed''
case of figure 2.
Initially, each
impulse causes every $x_j$ to increment by the same amount, so that
only the spring near the nail stretches.
As time progresses, this spring stretches more and more, until the
force it exerts becomes large enough to keep the first particle
from going into the next well under application of $A_1$,
the impulse with the smaller fractional part.
(At this point, the spring
force is insufficient to change the action of $A_2$, the impulse with the
larger fractional part.)  The second spring then starts to stretch, which,
on the next iteration, gives just enough added force to restore the first
particle to its initial motion for {\em both} impulses.  So now
the first spring stretches on alternate
applications of the impulse $A_1$, and the total spring force
on particle 1 increments alternately by $+k$ and $-k$.
Therefore, $c_1$ oscillates around
the memory value $A_1$, leading to a plateau on the plot of figure 2.
Eventually the second spring is stretched to the point
that the second particle also hangs up at the impulse $A_1$.
This, in turn, starts the
stretching of the third spring, etc.
A memory for $A_1$ is created whenever
the local curvature just cancels the fractional part of $A_1$.
A similar
analysis holds for all the other impulses $A_\tau$ that are applied.  As
time progresses, the $c$'s get stuck at all the different possible
memory values.

Another way to understand why the curvatures stick at the values
of $A_\tau$ is to note that
when $c_j$ passes through a memory value, then the right hand
side of Eq.~(\ref{eq:curveq}) changes discontinuously, and in
particular can change sign if the neighboring $c$'s have the
appropriate values.
If this happens, then $c_j$ oscillates with
amplitude $\propto k$, and sticks
at the memory value.\cite{tangnote}
This sticking can take the form of
either a fixed point or a
cycle in the local $c$-values.  

For $k \rightarrow 0$, the dynamics
separates into three regimes.
The smallest motions are the $O(k)$ back-and-forth motions
at the memory values, which lead to minute serrations in the
plateaus that are not visible on the scale of our figures.
The largest motions, involving changes in $c_j$ which are larger
than of order unity, and changes in $\tau$ which are much larger
than $1/k$,
are described by a linear discrete diffusion equation
\begin{equation}
\tilde{c}_j(\tau +1) - \tilde{c}_j(\tau) = k
\sum_{i~(nn)} (\tilde{c}_i(\tau)-\tilde{c}_j(\tau)) \ ,
\label{eq:linmap}
\end{equation}
obtained by
linearizing
the ${\rm int}$ functions in Eq.~(\ref{eq:curveq}).
Figure 4, which shows snapshots of configurations
at two different times,
demonstrates the accuracy of the linearized equation
in reproducing the evolution on large scales.
Numerically, the maximum deviation between the solutions
to Eqs.~(\ref{eq:curveq}) and (\ref{eq:linmap}) for identical
initial conditions,
$\sup_{j,\tau} |c_j(\tau)-\tilde{c}_j(\tau)|$,
is less than unity
for all system sizes, parameter values, boundary conditions, and initial
conditions investigated.
Using the nature of the nonlinearity in Eq.~(\ref{eq:curveq})
together with the fact that $|z-{\rm int}(z)| \le 1$ for all $z$,
one can obtain a rigorous analytic bound on this difference
that grows as $\log(L)$, where $L$ is the system's linear
extent,\cite{lpk_unp}
which is sufficient to insure the applicability of the
memory duration estimates given below.

Of course, neither of the regions just discussed
produces the memory effects.
The memories
come from an intermediate region involving variations in $c_j$ much
larger than $k$ but smaller than unity.  Figure~2 shows that, on
this intermediate scale, all the $c_j$'s show a very simple
behavior. They are either
(1) stuck at one of the memory values, or
(2) are between memory values, with the $c_j$ varying linearly in
time.
This sequence of step-wise linear motions
progressively reduces
the variation in
the discrete curvature of $c_j$.
Thus, the intermediate and large scale
motions of $c_j$ can be described as different kinds of diffusive
smoothing.

To characterize how the memory durations depend on system parameters,
first consider the case when the number of memories $M=1$.
The onset time $\tau _{onset}$ is determined by the condition
that the {\em curvature of the curvature} $\nabla^2 c$ is $\sim 1$.
This condition follows because the $c$'s stick only when the
discontinuity in the integer function is large enough to cause the
right hand side (rhs) of Eq.~(\ref{eq:curveq}) to change sign.
For multiple memories, we sum the rhs
of Eq.~(\ref{eq:curveq}) over the $M$ terms in the cycle and
note that the discontinuity occurs in only one of $M$ terms in
the sum, leading to the condition $\nabla^2 c \sim 1/M$.
For any $M$,
the transient memories disappear when the {\em range}
of curvatures becomes too small,
$\delta c \equiv c_{max}-c_{min} \sim 1$.
To relate the onset and forgetting times to these conditions
on the curvatures, we use the linearized map Eq.~(\ref{eq:linmap}),
whose evolution obeys (when $\tau \gg 1/k$):
\begin{equation}
c_j(\tau) = \sum_{q} e^{i\vec{q} \cdot \vec{j}}c(q,\tau =0)
e^{-kq^2 \tau }\ ,
\end{equation}
where $c(q,\tau =0)$ is the spatial Fourier transform of the initial
condition $c_j(\tau =0)$.
For the random initial conditions for $c$ shown in figure 3,
$\nabla^2 c$ is dominated by $q \sim \sqrt{k\tau}$ and
the onset time $\tau _{onset} \sim c_0M/k$,
where $c_0$ is a typical value of $c(q, \tau=0)$, independent of
the linear extent L.
At long times $\delta c$ is dominated by the longest wavelength mode,
so that $\tau _{forget} \sim \frac{L^2}{4\pi^2k}\ln (c_0)$.
Thus, larger systems remember longer.

The system of Eqs.~(\ref{eq:curveq}) is deterministically driven towards a
fixed point.
Once this point is reached, it is impossible to retrieve the short-term
memories.  However, it is possible to keep the transient memories from
decaying by adding noise to the system.\cite{pov_unp}
For example, in the ``nailed''
case of Figure 2, moving the nail slowly but randomly with time creates the
possibility of continuously encoding new memories.  That noise can lead to
the {\em retention} of memories is important for understanding our
experiments.  Permanently encoded multiple memories are observed,
but only in samples of $NbSe_3$ with additional
conducting strips attached to the crystal between the contacts,
an arrangement known to induce noise.\cite{thorne_noise}

Our memory mechanism can be compared to the ``Hopfield
memory,''\cite{neuralnetrefs} which is a dynamical system with parameters
adjusted so that particular configurations, which encode the desired
patterns, minimize an energy functional.  There, the memory is encoded in
the long-time dynamics, and there is no intrinsic ``forgetting'' mechanism.
Moreover, changing the remembered value requires nontrivial adjustment of
the microscopic couplings of the model.  In the CDW system studied here,
the information is encoded in an evolving system,
the control parameter (pulse size) is easily varied in the laboratory, and
the self-organization is exhibited via standard transport measurements.

%
One
avenue for further investigation is to characterize better the effects
of noise, which, as discussed above, plays an important role in the CDW
experiments.
We would also like to identify other
experimental systems which exhibit this short-term memory.
Because the present model
is just a discretized diffusion equation,
having properties which seem quite robust,  
we are optimistic that physical embodiments exist that could be made
into useful devices.

We thank S.E. Brown, P.B. Littlewood, M.L. Povinelli,
and R. Thorne for fruitful
discussions.  This work was supported in part by the MRSEC program
of the National Science Foundation under award number
DMR-9400379.
A.L. acknowledges support by an AT\&T Graduate Fellowship.


\begin{figure}
\epsfxsize = \hsize
\centerline{\epsfbox{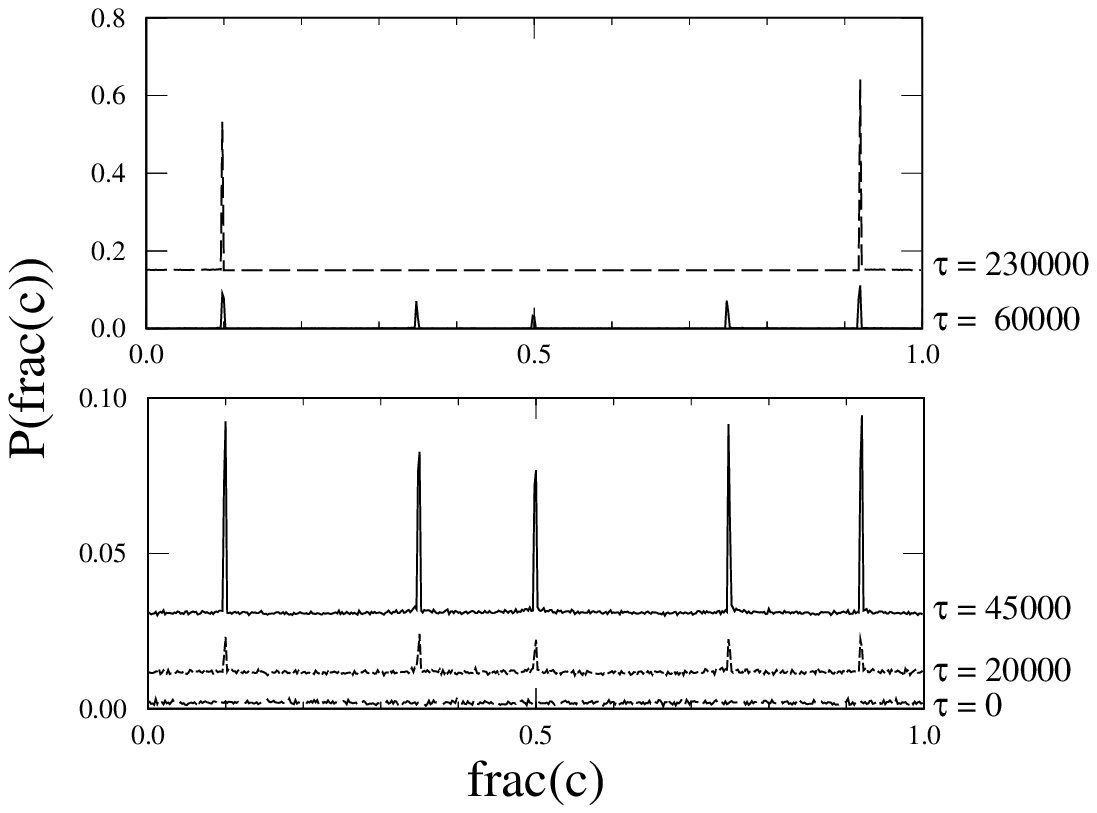}}
\vspace {1cm}
\end{figure}
\figure{Figure~1: Memory formation of Eq.~(\protect\ref{eq:xeq}) in a
two-dimensional $100 \times 100$ system with periodic boundary conditions,
$k=0.0001$, initial condition of $x$'s chosen randomly from the interval
$[0,1000000]$, and $M=5$ ($A = [0.1,0.35,0.5,0.75,0.92]$).  $P$ is the
proportion of $c$'s with fractional parts {\rm frac}(c) within a bin of
width 0.002.  For clarity, successive curves are offset vertically.  The
lower panel illustrates the short-term accumulation of $c$'s at each value
of $A$; the upper panel demonstrates that at long times only two peaks
persist.
\label{fig1}}

\begin{figure}
\epsfxsize = \hsize
\centerline{\epsfbox{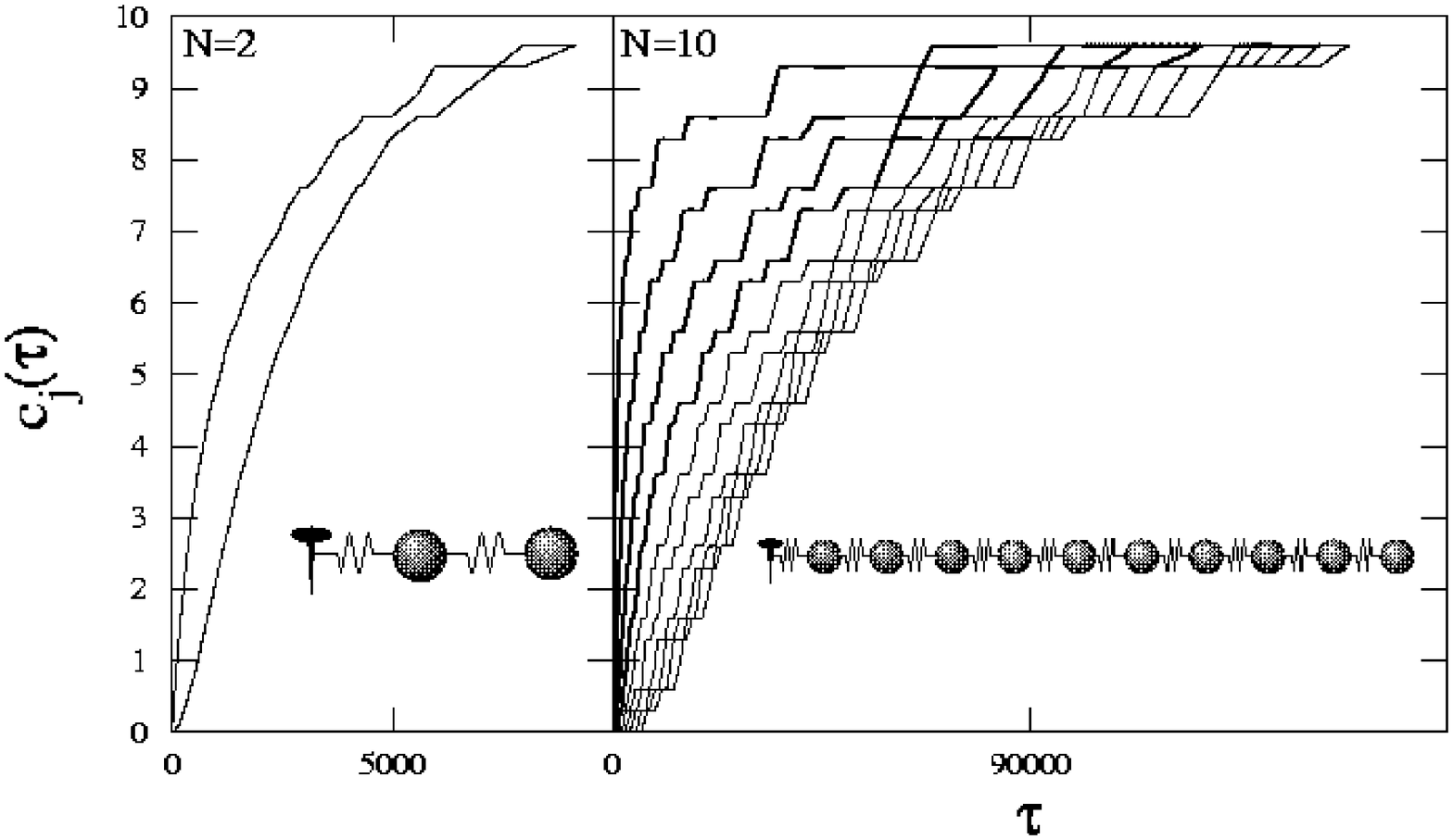}}
\end{figure}
\figure{Figure~2:
Plot of curvatures $c_j(\tau)$ versus $\tau $ for $N$-particle chains with
fixed-end boundary conditions Eq.~(\protect\ref{eq:fixedendbc}), $k=0.001$
and $A = [9.3,9.6]$, and initial condition $x_j(\tau=0) = 0$ for all $j$.
The memories are manifest in the plateaus (more pronounced for $N=10$ than
for $N=2$) when the $c_j$ have values with fractional part of 0.3 and 0.6.
Only one memory is retained at long times.
\label{fig2}}

\begin{figure}
\epsfxsize = \hsize
\centerline{\epsfbox{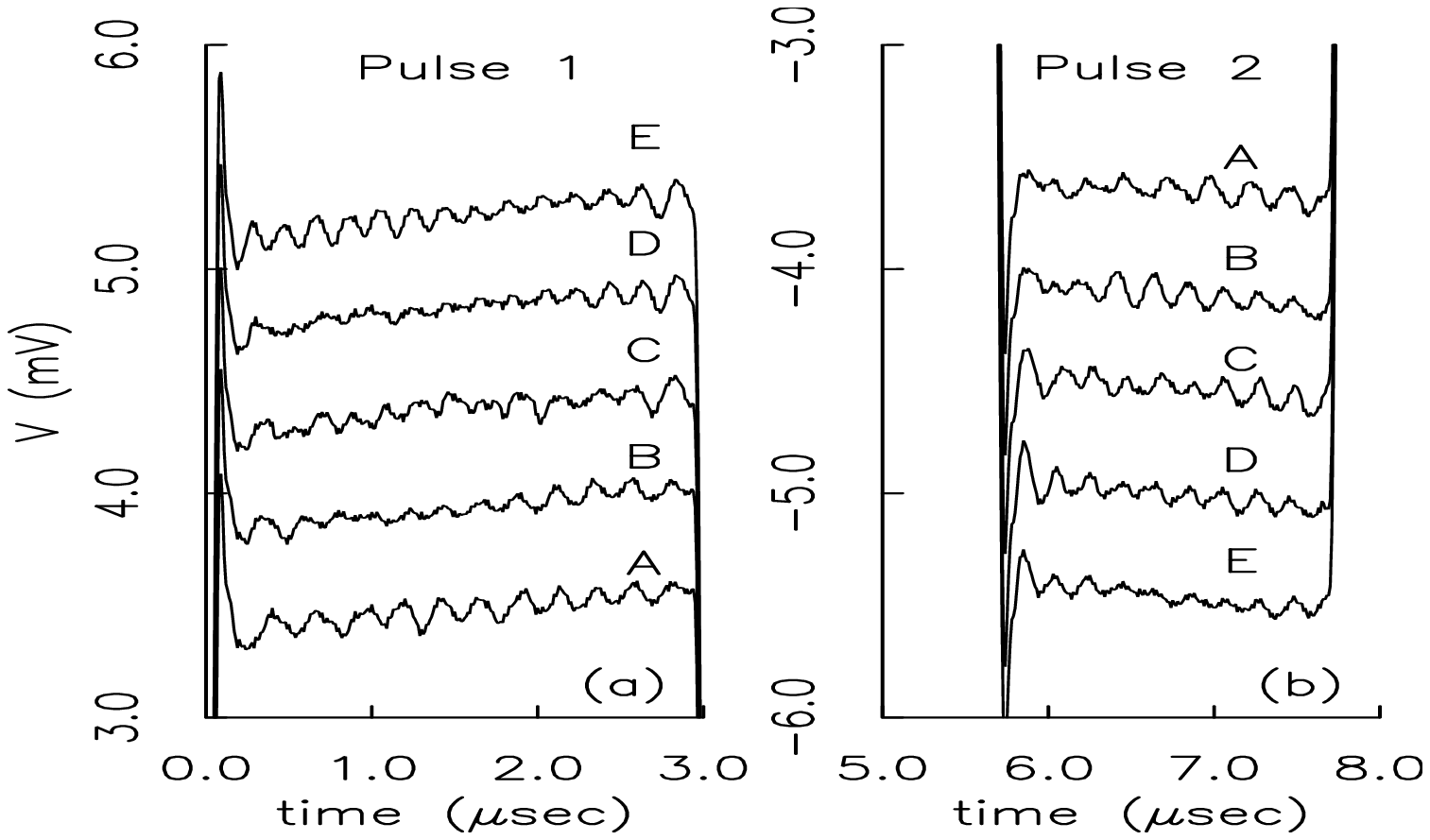}}
\end{figure}
\figure{Figure~3:
Trained voltage response curves of $NbSe_3$ for five different four-pulse
sequences.  The evidence for multiple-memory encoding is the decreasing
magnitude of the voltage at the end of each pulse. (Note the two pulses
have opposite signs. The voltage response for pulses 3 and 4 are not
shown.)  The drive is a sequence of 4 current pulses with magnitudes $[I,
-I, I, -I]$ and durations $2.95$, $2.05$, $1.75,$ and $2.75$ $\mu s$. The
drive magnitude, $I$, is 20.76, 21.16, 21.55, 21.94, and 22.33 $\mu A$ for
curves A, B, C, D, and E, respectively.  Training consisted of over one
million pulse repetitions.  Measurements were performed using a two-wire,
silver-paint contact configuration.  Sample dimensions are 5.2 $\mu m^2
 \times $ 980 $\mu m$.  Additional silver paint strips of
43 $\mu m$ and 100 $\mu m$ in width are attached to the sample, centered at
distances 13\% and 58\% between the probe contacts respectively.  The
curves are offset for clarity, and averaged 200 times to reduce noise.
$T=50K$, and $E_T=47~mV/cm$.
\label{fig3}}

\begin{figure}
\epsfxsize = \hsize
\centerline{\epsfbox{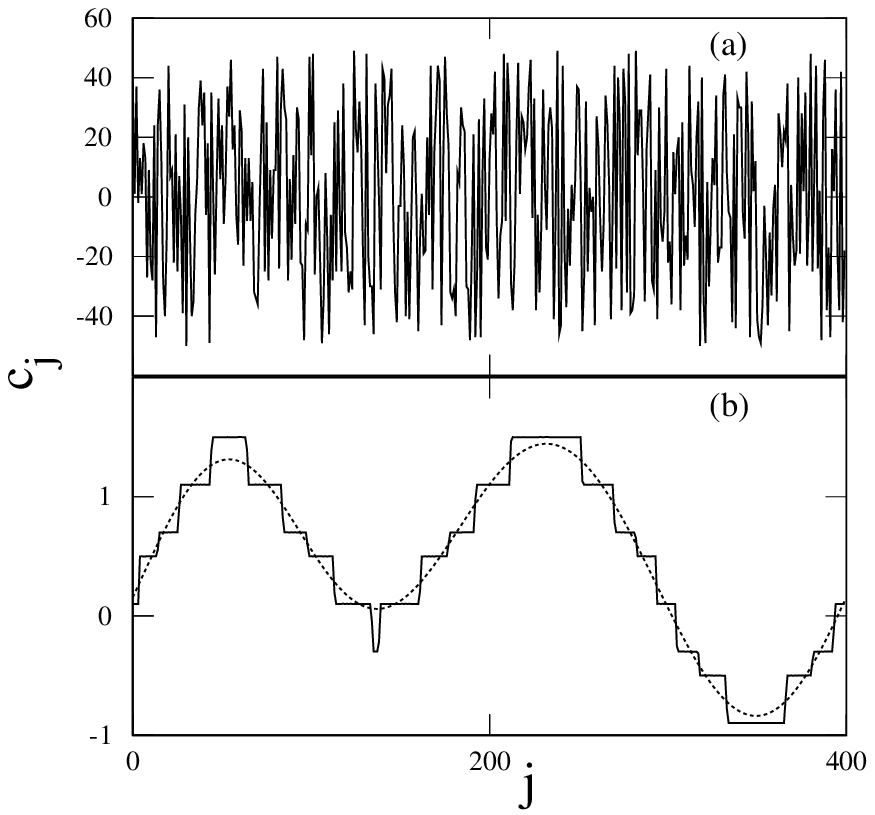}}
\end{figure}
\figure{Figure~4:
Snapshots of curvature $c_j$ versus position $j$ of a one-dimensional chain
of length L=400 with periodic boundary conditions at (a) $\tau=0$
and (b) $\tau=1200000$.  (The fixed point is reached at
$\tau \sim 9300000$.)  Parameter values are $A=[1.1, 1.5, 1.7]$, $k=0.001$.
The dashed line in (b) is the configuration obtained using
Eq.~(\protect\ref{eq:linmap}) with the same initial conditions,
demonstrating that this linear diffusion equation captures accurately
the large scale evolution of the system.
\label{fig4}}


\begin{references}
\bibitem{expfootnote} Multiple memory encoding has been observed
in two out of three samples of $NbSe_3$ which have additional
conducting strips attached to the crystal between the contacts.
The role of these extra strips is discussed below.
\bibitem{tangnote}  Ref.~\cite{tang87} notes the importance of the
		oscillations induced by the nonlinearity in the
		context of the fixed point selection when a single
		memory is encoded.
\end{references}
\end{document}